\documentclass[
 reprint,
 amsmath,amssymb,
 aps,
]{revtex4-2}
\usepackage{xcolor}
\usepackage{graphicx}% Include figure files
\usepackage{dcolumn}% Align table columns on decimal point
\usepackage{bm}% bold math
\usepackage{booktabs}
\begin{document}

\preprint{APS/123-QED}

\title{A method for reversing the laser modulation in a storage ring}% Force line breaks with \\

\author{Weihang Liu}

\affiliation{Institute of High Energy Physics, Chinese Academy of Sciences, Beijing 100049, China}%
\affiliation{ Spallation Neutron Source Science Center, Dongguan 523803, China}%

\author{Yu Zhao}%

\affiliation{Institute of High Energy Physics, Chinese Academy of Sciences, Beijing 100049, China}%
\affiliation{ Spallation Neutron Source Science Center, Dongguan 523803, China}%

\author{Yi Jiao}%
 \email{ jiaoyi@ihep.ac.cn}

\affiliation{Institute of High Energy Physics, Chinese Academy of Sciences, Beijing 100049, China}%

\author{Xiao Li}%
\affiliation{Institute of High Energy Physics, Chinese Academy of Sciences, Beijing 100049, China}%
\affiliation{ Spallation Neutron Source Science Center, Dongguan 523803, China}%

\author{Sheng Wang}%
 \email{ wangs@ihep.ac.cn}
\affiliation{Institute of High Energy Physics, Chinese Academy of Sciences, Beijing 100049, China}%
\affiliation{ Spallation Neutron Source Science Center, Dongguan 523803, China}%

\author{Chao Feng}%
\affiliation{Shanghai Institute of Applied Physics, Chinese Academy of Sciences, Shanghai 201800, China}%
\affiliation{Shanghai Advanced Research Institute, Chinese Academy of Sciences, Shanghai 201210, China}%

\begin{abstract}
The pursuit of coherent radiation generation remains a key direction in the advancement of storage ring light sources. Despite the potential of laser modulation in achieving this goal, it leads to a significant decline in the quality of the electron beam. Efforts to mitigate this decline have resulted in the proposal of demodulation schemes. However, implementing modulation and demodulation within the storage ring presents significant challenges due to dynamical and spatial constraints within straight sections. In this study, we propose a straightforward and easily implementable method for achieving reversible laser modulation in a storage ring. Notably, our approach circumvents the need for lengthy straight sections or bypass section. Simulation results show a substantial restoration of beam quality following demodulation. This innovative scheme holds great promise for the realization of high repetition rate coherent storage ring light sources.

\end{abstract}

\maketitle

%\tableofcontents

\section{\label{sec:level1}INTRODUCTION}
As a crucial experimental tool for scientific inquiry, the performance of light sources continues to evolve in tandem with scientific progress. Notably, particle accelerator-based light sources are continually advancing their technology and capabilities to meet the demands of scientific exploration. The free electron laser (FEL) exemplifies this progression, transitioning from normal conducting accelerators to superconducting accelerators. This upgrade has significantly increased the repetition rate of radiation pulses by nearly five orders of magnitude \cite{decking_mhz-repetition-rate_2020}. Furthermore, leveraging sophisticated beam manipulation techniques, FELs now offer temporal resolutions on the scale of attoseconds \cite{duris_tunable_2020}, empowering attosecond science with an unparalleled tool \cite{li_hearing_nodate}.

Similarly, the storage ring light source (SRLS) has also undergone significant enhancements. By adopting the multibend achromat (MBA) design \cite{einfeld_first_2014}, SRLS has achieved a substantial reduction in its natural emittance, approaching the diffraction limit of X-rays. Consequently, the brightness of the light source has surged by more than two orders of magnitude, heralding a new era of enhanced performance and capabilities.

The high repetition rate, multi-user capability and high stability of SRLS make it more attractive than FEL. However, its longitudinal coherence and time resolution are still not comparable to FELs. At present, the time resolution of SRLS is basically in the picosecond range or longer. In addition, its energy resolution is also limited by the monochromator technology, which is difficult to break through to the millielectronvolt scale. These limitations have made it difficult to meet some of the needs of current and future scientific research, and it is therefore necessary to be improved.

The generation of microbunches via laser modulation represents a viable solution for fulfilling this requirement. By interacting the beam with a laser in a modulator, the beam is given an energy modulation. This energy modulation is converted to density modulation after the beam passes through a longitudinal dispersion section. This density modulated beam will generate coherent radiation in the downstream undulator at high harmonic of the laser \cite{girard_optical_1984,de_ninno_generation_2008,evain_soft_2012,liu_generating_2018,yang_optimization_2022,hwang_generation_2020}. Due to the properties of coherent radiation, the radiated power is proportional to the square of the beam charge and the pulse duration is proportional to the modulated laser pulse duration. These properties are capable of generating radiation pulses with high power and variable duration. 

By combining the "flat" nature of the storage ring, i.e., its ultra-low vertical emittance, numerous laser modulation schemes better suited for the storage ring have been proposed. For instance, vertical momentum (or angular) modulation can be induced in the beam using a TEM01 or tilted incidence laser \cite{zholents_attosecond_2008,xiang_generating_2010,wang_angular_2019,lu_methods_2022}. This modulation can then be converted into density modulation as the beam traverses through a dispersion section, benefiting from the ultra-low vertical emittance, which significantly enhances the density modulation. Furthermore, energy modulation schemes can also benefit from the ultra-low vertical emittance. One such method, called angular dispersion microbunching generation (ADM) \cite{feng_storage_2017}, utilizes a dipole to couple the longitudinal and transverse motion of an electron prior to energy modulation. The resultant energy modulation is then converted into density modulation through a dogleg. This approach has been utilized in the design of a coherent radiation beamline within a straight section of a storage ring  \cite{li_extremely_2020,liu_generating_2023}.

Regardless of the type of modulation, it inevitably affects the beam, leading to a degradation in beam quality. This degradation makes it challenging to modulate the beam continuously or at high frequencies, thus limiting the repetition rate. To address this issue, demodulation schemes have been proposed \cite{jiang_synchrotron-based_2022,li_generalized_2023}. These schemes restore beam quality by demodulating the modulated beam.

The demodulation process for energy modulation is relatively straightforward compared to angular modulation, as it primarily involves removing the energy modulation. In contrast, angular modulation requires additional considerations, such as betatron motion in the vertical plane and precise matching of the beta function phase shift. Therefore, choosing an ADM modulation scheme makes it easier to design a demodulation transport. However, considering the complexity of the storage ring dynamics and the limited space of the straight sections, it is extremely challenging to design a reversible ADM within a single straight section of a storage ring. The main difficulty is to design an isochronous transport between energy modulation and demodulation. This requires sufficient space on the one hand and a delicate design on the other to avoid excessive impact on the storage ring dynamics.

There are two ways to handle this difficulty, one is by designing a dedicated long straight section storage ring \cite{li_generalized_2023}. The other is to place the reversible ADM in the bypass of the storage ring \cite{jiang_synchrotron-based_2022}. The bypass design requires frequent injection and extraction to achieve coherent radiation, and the technical constraints of the injection and extraction elements impose limitations on the achievable repetition rate. On the other hand, designing a dedicated long straight section is not a standard requirement for a typical SRLS. Additionally, the layout of the bypass and the long straight section can be designed to support a ring-based FEL \cite{MURPHY1985159,KIM198554,NUHN199289,Lindberg2013,Cai01052013,Mitri_2015,HUANG2008120,DIMTR060702}, which could make the laser modulation-based coherent radiation methods less attractive.

To enable the implementation of a reversible ADM in a typical storage ring, this paper presents a simple scheme. In this approach, the ADM and the demodulation section are situated in distinct straight sections, while an isochronous transport is established using arc sections and a horizontal chicane positioned between the ADM and the demodulation sections. The feasibility of the scheme is verified through simulations based on the parameters of the southern advanced photon source (SAPS) \cite{zhao_design_2021}. Post-modulation, a micro-bunch structure is achievable, demonstrating sufficient capability for producing coherent radiation. Following demodulation, beam parameters are largely restored. After the beam passes the demodulation section and completes the ADM process, the lasers are turned off, allowing the beam to undergo damping in the ring for several turns. This process is followed by reactivating the lasers, thus repeating the cycle. Simulations suggest that approximately 600 turns of damping are necessary to achieve stable and optimal radiation performance. This means that coherent pulses appear in a burst mode. As shown in Fig. 1, a possible burst mode is illustrated. In this mode, coherent pulses are not generated uniformly but rather as a series containing the same number of pulses as the bunches. This coherent pulse train repeats every 600 turns, equivalent to a repetition rate of 0.25 MHz. The resulting average coherent radiation flux can reach $10^{15}$ photons/s, with a bandwidth of less than $10^{-4}$. In contrast, under similar bandwidth conditions, the photon flux from synchrotron radiation is constrained to approximately $2.5\times 10^{13}$ photons/s, indicating a reduction of more than one order of magnitude compared to coherent radiation. This substantial enhancement underscores the potential of the proposed method for advancing new directions in the development of SRLS.

In the following sections, we will present detailed information regarding the proposed method, including a specific example demonstrated on SAPS in Section 3. The repetition rate and radiation performance will be discussed in Section 4. Finally, based on these results, we will provide our discussion and conclusions in Section 5.

\section{Reversible ADM Implemented in Straight Sections of a Storage Ring}

To enhance the utilization of a storage ring for a given circumference, it is common practice to limit the length of straight sections to a relatively short range, typically 5-6 meters. In order to achieve reversible ADM in such storage rings, it is necessary to place modulation and demodulation in separate straight sections. Typically, storage rings are multi-period structures; for instance, the SAPS has a period of 32. In this configuration, the ADM section and the demodulation section are positioned in two adjacent periods, with a horizontal chicane situated between them in the straight section, as illustrated in Fig. 1.

\begin{figure*}
    \centering
    \includegraphics[width=0.8\linewidth]{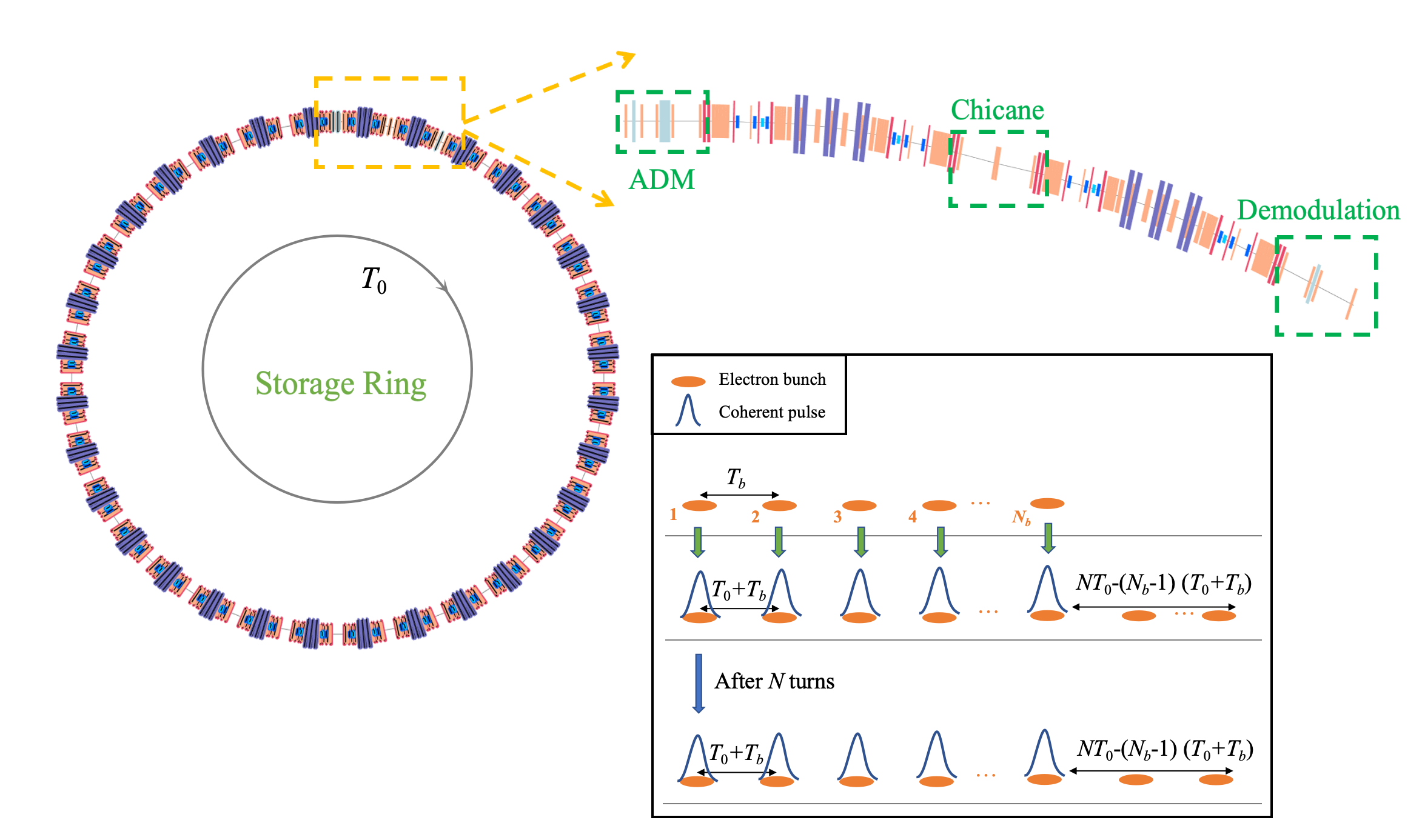}
    \caption{Layout of a reversible ADM in a storage ring with a potential burst mode. In this mode, the storage ring accommodates $N_b$ bunches. After the first bunch is modulated, the initial coherent pulse is generated, followed by sequential modulation of the remaining bunches at intervals of $T_0 + T_b$. Once all $N_b$ bunches are modulated, the cycle restarts from the first bunch if it has completed $N$ turns and its beam parameters are restored. Otherwise, the system waits for $NT_0 - (N_b - 1)(T_0 + T_b)$ before resuming modulation.}
    \label{fig:1}
\end{figure*}

This layout enables the achievement of isochronous transport from the modulation section to the demodulation section. Following energy modulation in the ADM section, the beam proceeds through downstream components including a dogleg, radiator, and dipoles, contributing to the longitudinal dispersion $R_{56,\mathrm{adm}}$ in the isochronous transport. Subsequently, the beam traverses the first arc region after the ADM section to reach the horizontal chicane. It then passes through the second arc region to access the demodulation section. Within the demodulation section, the beam encounters two vertical dipoles before reaching the modulator, which also imparts longitudinal dispersion $R_{56,\mathrm{dem}}$. Summing all longitudinal dispersion contributions yields the total longitudinal dispersion of the isochronous transport, $R_{56,\mathrm{tot}} = R_{56,\mathrm{adm}} + 2R_{56,\mathrm{r}} + R_{56,\mathrm{c}} + R_{56,\mathrm{dem}}$, where $R_{56,\mathrm{r}}$ and $R_{56,\mathrm{c}}$ represent the arc and chicane longitudinal dispersion, respectively. Following the isochronous requirement, we obtain

\begin{equation}
\begin{cases}
R_{53,\mathrm{tot}}=0\\R_{54,\mathrm{tot}}=0.\\R_{56,\mathrm{tot}}=0
\end{cases}
 \label{eq:1}
\end{equation}

The achievement of zero $R_{56,\mathrm{tot}}$ benefits from the fact that the longitudinal dispersion $R_{56,\mathrm{r}}$ of a typical storage ring arc section is opposite in sign to the other contributing terms of $R_{56,\mathrm{tot}}$. 

Given the ADM parameters, the isochronous requirement can be achieved by adjusting the dipole angle $\theta$ of the chicane dipoles and the angles $\theta_4$ and $\theta_5$ of the two dipoles upstream of the modulator in the demodulation section. It's worth noting that both the modulation and demodulation sections are located in straight sections, so $R_{51,\mathrm{tot}}$ and $R_{52,\mathrm{tot}}$ naturally become zero.

The ADM section needs to be designed to target the coherent radiation performance or the magnitude of the bunching factor for a particular harmonic. The ADM bunching factor is related to the beam parameters as \cite{feng_storage_2017}:
\begin{equation}
b(n)=J_{n}\bigl(nk_{L}r_{56}A\bigr)\exp[-\frac{1}{2}(nk_{L}\eta)^{2}\gamma_{y}\epsilon_{y}],
 \label{eq:2}
\end{equation}
where $r_{56}$ and $\eta$ are the dogleg vertical and longitudinal dispersions of the ADM, respectively, $A$ is the modulation strength, $\gamma_y$ is the twiss parameter of the ADM entrance in the vertical direction, $\epsilon_y$ is the vertical emittance, and $k_L$ is the wave number of the laser.

Equation (2) indicates that the storage ring vertical emittance should be controlled at a relatively low level to obtain a sufficiently large bunching factor. Since both the ADM section and the demodulation section contribute to the vertical dispersion, the dipoles, modulators, and radiator in both sections contribute to the quantum excitation, increasing the vertical emittance:
\begin{equation}
\epsilon_y\propto\int_{\mathrm{dipole,mod,rad}}\frac{H_y}{\mid\rho\mid^3}ds,
 \label{eq:3}
\end{equation}
where $H_y$ is the vertical chromatic $H$-function. The most effective way to reduce the growth of vertical emittance is to minimize the dispersion of these two sections.

The layout of the ADM section, as depicted in Fig. 2, consists of five dipoles, a modulator, and a radiator. To ensure achromatic conditions and coaxiality of the entrance and exit beams, the deflection angle and spacing of the dipoles must satisfy the following \cite{zeng_asymmetric_2024}:
\begin{widetext}
\begin{equation}
\begin{cases}b( L_1+ L_2+L_3+ L_4+\frac{9 L_b}{2})+( L_2+ L_b)\theta_1+\biggl( L_4+\frac{3}{2} L_b\biggr)\theta_2+\frac{1}{2} L_b\theta_3=0\\{}\\b+\theta_2+\theta_3=0\end{cases}.
 \label{eq:4}
\end{equation}
\end{widetext}

Here, $b, \theta_1, -\theta_1, \theta_2, \theta_3$ correspond to the deflection angles of the first to the fifth dipole, $L_i$ ($i=1,2,3,4$) represents the distance between the dipoles, and $L_b$ represents the length of each dipole.

Under these conditions, combined with the ADM optimality condition \cite{feng_storage_2017}
\begin{equation}
\left.\left\{\begin{array}{l}r_{56}+b\eta=0\\\\1+Ak_{L}r_{56}=0\end{array}.\right.\right.
 \label{eq:5}
\end{equation}

It can be observed that the dispersion of the ADM section is positively correlated with the deflection angle $b$, while the bunching factor is also positively correlated with $b$. Therefore, there is a trade-off between low dispersion and a high bunching factor. Given the modulation strength, one can optimize the bunching factor to find deflection angle $b$ or to choose a small enough one.

\begin{figure}[h]
    \centering
    \includegraphics[width=1\linewidth]{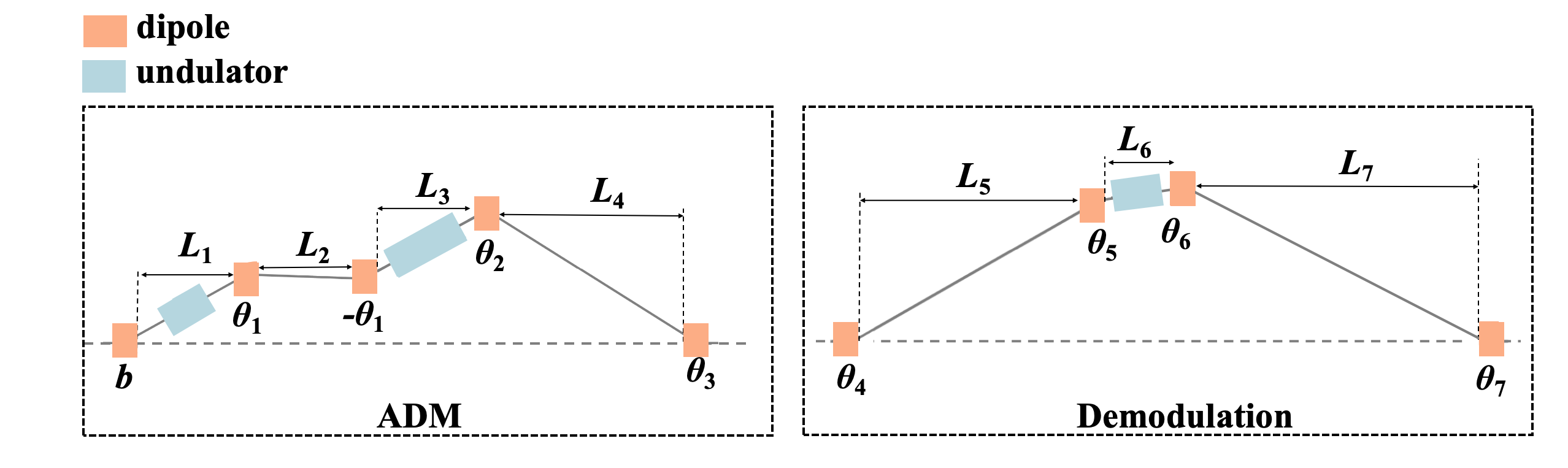}
    \caption{Layout of the ADM and Demodulation in straight sections.}
    \label{fig:2}
\end{figure}

The demodulation section as shown in Fig. 2, has a similar layout of the ADM section by comprises only four dipoles and a modulator. Once the parameters of the ADM section are defined, the parameters of the two dipoles preceding the modulator are exclusively determined by the isochronous condition [Eq. (1)]. Similarly, the parameters of the last two dipoles are determined by the achromatic and coaxial conditions.

The method described above facilitates the implementation of reversible ADM designs for typical storage rings. This approach is easier to construct and optimize compared to the scheme based on the lengthy straight section \cite{li_generalized_2023}.

\section{Reversible ADM based on SAPS}
To demonstrate the proposed scheme, we utilize the SAPS storage ring for reversible ADM design. It is worth noting that while this design may not be optimal, it sufficiently showcases the proposed approach. The SAPS storage ring operates at an energy of 3.5 GeV with a natural emittance of 33 pm and a straight section length of 6 m. 
\subsection{ADM and Demodulation design}

In the ADM section, we adopt an energy modulation intensity of 0.14\% and a modulated laser wavelength of 266 nm. In practical operation, we can fine-tune this wavelength and then adjust the dipole strength in the ADM, chicane, and demodulation sections to match the ADM and demodulation conditions. Additionally, this high repetition rate and high-power laser may be generated by the high average power fiber laser \cite{Fu21}, which could potentially produce lasers with average power at the hundred kW level \cite{Muller20}.

Somewhat arbitrarily, the deflection angle of the first dipole in the ADM section is set to 5 mrad. Given the limited space available in the straight sections, the modulator has a period number of 1 with a length of 0.25 m. The radiator has a period number of 7 with a length of 0.84 m. With these parameters, the maximum dispersion of the ADM section is calculated to be approximately 11 mm, as illustrated in Fig. 3. 

\begin{figure}[h]
    \centering
    \includegraphics[width=1\linewidth]{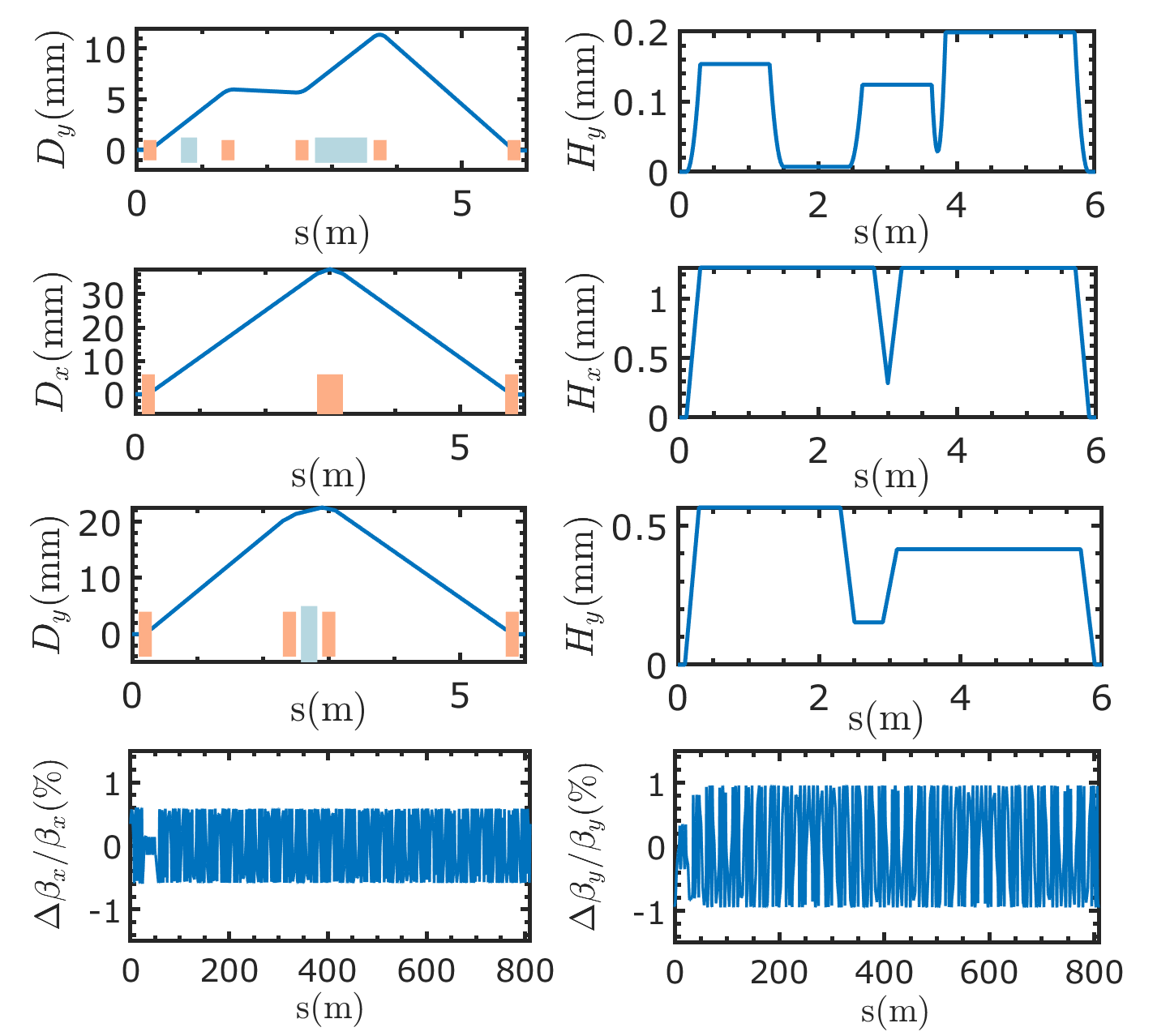}
    \caption{Dispersion and \textit{H}-function of the ADM (first row), Chicane (second row) and demodulation (third row) sections and beta-beating of the storage ring (fourth row).}
    \label{fig:3}
\end{figure}

To achieve isochronous transport between the modulation and the demodulation, the chicane parameter $\theta$ and the parameters $\theta_4$, $\theta_5$ of the vertical dipoles upstream of the modulator in demodulation section are set to 13.9 mrad, 9.59 mrad, and -6.73 mrad, respectively. Other parameters are detailed in the Appendix. These specified parameters result in a horizontal dispersion of 37 mm. Additionally, the demodulation section contributes a vertical dispersion of 22.5 mm, as shown in Fig. 3. 

The beta-beating, which represents the change in the beta function between the ring with and without the reversible ADM, is also shown in Fig. 3. The results demonstrate that the beta-beating is smaller than 1\%. It is important to recognize that such a low beta-beating has only a minimal impact on the overall beta-beating caused by other IDs with reasonable errors and misalignments. Therefore, it is possible to correct the beta-beating of the ring to an acceptable level. However, specifically for our work, we found that this low beta-beating has little impact on the results, so we chose not to implement this correction.

After completing the design, we calculated the dynamic aperture (DA) and the local momentum apertures (MA). As shown in Fig. 4, after incorporating the ADM, chicane, and demodulation sections, the DA area decreased by only 10\%, with the momentum aperture experiencing the largest reduction in the chicane section, dropping to 4\%. However, in the ADM and demodulation sections, the MA remains above 4\%. These changes have a negligible impact on the routine operation of the storage ring.

\begin{figure}[h]
    \centering
    \includegraphics[width=1.0\linewidth]{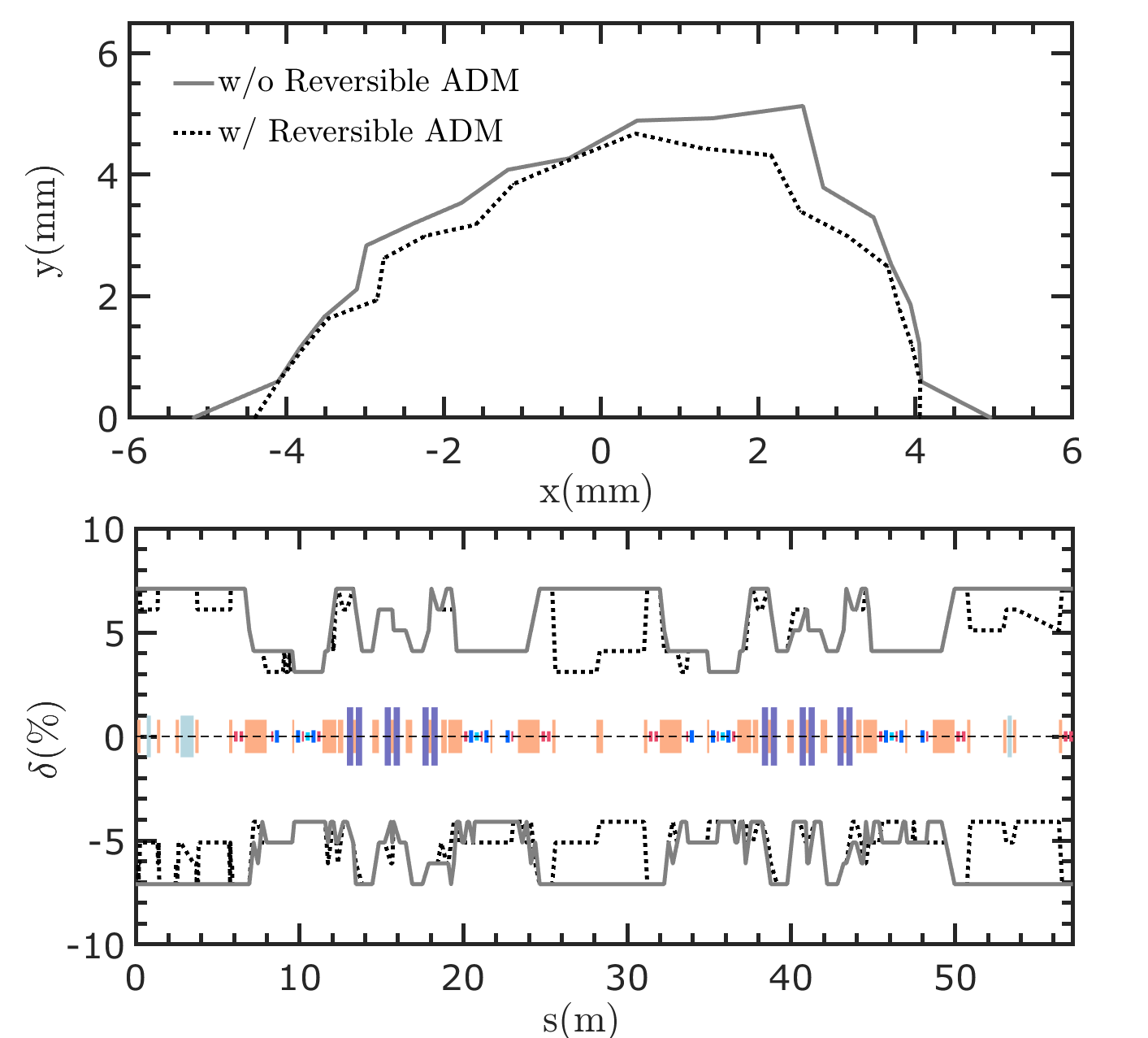}
    \caption{Dynamic aperture and local momentum apertures for SAPS ring with and without ADM, chicane and demodulation sections.}
    \label{fig:4}
\end{figure}

The final emittances are 36.82 pm in the horizontal direction and 1.2 pm in the vertical direction. We adopted the ELEGANT \cite{borland_elegant_2000} code to implement laser modulation and tracking the modulated beam through the ADM section, arc, chicane and the demodulation section. After the density modulation, the bunching factor versus the harmonic number is illustrated in Fig. 5. It can be observed that the 10th harmonic bunching factor is up to 20\%.
\begin{figure}
    \centering
    \includegraphics[width=0.8\linewidth]{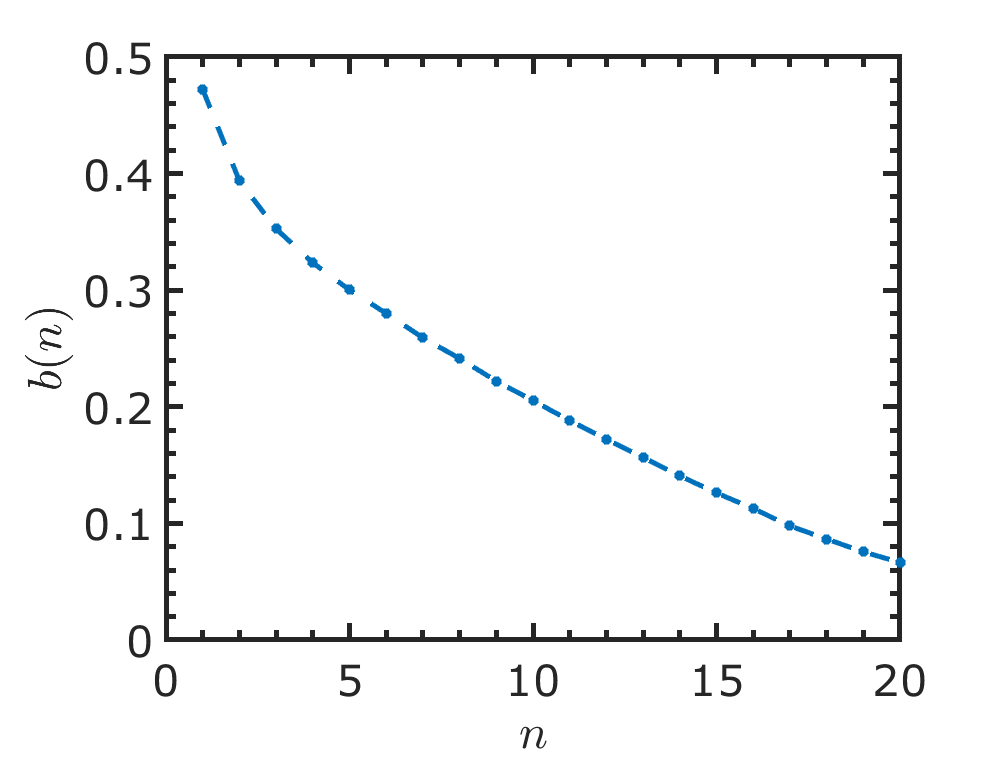}
    \caption{Bunching factor at different harmonic number.}
    \label{fig:5}
\end{figure}

During the demodulation process, perfect synchronization between the modulated and demodulated lasers is assumed. The detailed implementation of the optical path will be explored in future studies. Upon completion of the demodulation, the beam's phase space is effectively restored. This is attributed to the achievement of isochronous transport, as shown in Fig. 6, where from the center of the ADM modulator to the center of the modulator in demodulation, $R_{5i,\mathrm{tot}} = 0$ ($i=$1,2,3,4,6). When considering only linear terms of the transport and treating the energy modulation as an equivalent sinusoidal point kick, following demodulation, the increase in energy spread and vertical emittance are nearly negligible. However, in a more realistic scenario that accounts for nonlinear terms of the transport, 3D laser-electron beam interaction, and finite-length modulators, the increase in energy spread rises to 0.08\%, and the increase in vertical emittance to 16.0\%, indicating a significant growth compared to the ideal case. This indicates a influence of nonlinear terms, and further reduction in parameters growth may be achievable by appropriately optimising the multipole magnets in the arc section. Nevertheless, this would likely compromise the symmetry of the ring and potentially affect its dynamics, addressing this issue may require different strategies and is beyond the scope of this paper. Fortunately, in our case, such optimization is not necessary.

\subsection{Intra-beam scattering effect and Touschek lifetime}

In the previous design, we only considered the case of zero beam current. When a certain beam current is considered, the intra-beam scattering (IBS) effect may become significant, as it further increases the emittance of the storage ring, thereby reducing the bunching factor of the ADM.

To analyze the IBS effect, we use Bane’s high-energy approximation \cite{BANE084403} to calculate the IBS diffusion times 
$T_x$,$T_y$, and $T_{\delta}$. It is important to note that the vertical emittance has two contributions: one from the vertical dispersion and the other from the coupling of the betatron motion. We define the factor of the betatron coupling as $\kappa_{\beta}$. Under the assumption of small $\kappa_{\beta}$, the beam's equilibrium parameters can be derived from the following equations \cite{BANE084403}:
\begin{equation}
    \begin{aligned}
& \epsilon_x=\frac{t_x}{T_x} \epsilon_x+\epsilon_{x 0} \\
& \epsilon_y=\kappa_\beta \epsilon_x\left(1-\frac{t_y}{T_y}\right)+\frac{t_y}{T_y} \epsilon_y+\epsilon_{y0,d} \\
& \sigma_\delta=\frac{t_\delta}{T_\delta} \sigma_\delta+\sigma_{\delta 0}
\end{aligned}
\end{equation}
Here $t_{x,y,z}$ are the synchrotron radiation damping times, and $\epsilon_{y0,d}$ represents the vertical emittance due to the dispersion contribution at zero beam current.

We consider the case where the single bunch charge is 1 nC and the betatron coupling is 0.1\%. Such a low coupling level is expected to be achievable through the resonance driving terms correction method \cite{FRANCHI034002}. To maintain emittance stability, we aim to limit the coupling variation to no more than 10\%, which would keep the jitter of the bunching factor and radiation power within 1\%.

We examine three different radiation wavelengths, with harmonic numbers $n$ of 8, 9, and 10. These different harmonics can be realized by adjusting the gap of the radiator. The beam parameters for these three cases are provided in Table 1.

\begin{table}[h]
\caption{Emittances for different harmonics with and without IBS}
\resizebox{\columnwidth}{!}{
\begin{tabular}{cccc}
\hline
Parameter & \multicolumn{3}{c} {value} \\
\hline
n & 8 & 9 & 10 \\
radiator peak field (T) & 0.63 & 0.59 & 0.54 \\
($\epsilon_x$, $\epsilon_y$) w/o coupling and IBS (pm) & (36.84, 1.27) & (36.83, 1.22) & (36.82, 1.20) \\
($\epsilon_x$, $\epsilon_y$) w/ coupling and IBS (pm) & (102.6, 1.95) & (103.2, 1.90) & (103.4, 1.88) \\
\hline
\end{tabular}
}
\end{table}

It is important to note that the emittance differences among these three cases are quite small, indicating that under constant betatron coupling, the effect of undulator focusing on the vertical emittance remains weak within the parameter range shown in Table 1. For larger gap variations, additional damping wigglers can be installed in the storage ring to stabilize the emittance \cite{TAN2023168278}.

For simplicity and without loss of generality, in the subsequent calculations, we assume the equilibrium emittance after considering IBS to be 103 pm horizontally and 2 pm vertically for all three cases presented in Table 1. The corresponding bunch length is 3.8 mm, the peak current is 31.5 A, and the energy spread is 0.14\%. Under these beam parameters, after demodulation, the energy spread increases by 0.1\%, and the vertical emittance increases by 22\%. This result is larger compared to the case without considering IBS, suggesting that the demodulation effect is dependent on the beam's initial parameters.

Based on the beam parameters above and the MA results provided in Section 3.1, we can estimate the Touschek lifetime of the demodulated beam. We use touschekLifetime tool in ELEGANT, which calculates the Touschek lifetime using A. Piwinski's formula \cite{piwinski1999}. The results show that the Touschek lifetime of the demodulated beam is approximately 1.44 hours. With this lifetime, to maintain a current stability of 0.1\%, a top-up injection mode can be employed, with injections occurring at a repetition rate of 0.1 Hz, which is achievable with the currently technology.

\section{Repetition rate and radiation performance}

The demodulated beam must complete an adequate number of turns in the storage ring (with the laser off) to restore its quality for the subsequent ADM process (with the laser on). It is crucial to avoid excessive turns, as this would lead to a reduction in the repetition rate. Conversely, if the number of turns is insufficient, the beam quality may fail to recover adequately, resulting in progressively poorer performance in the ensuing ADM processes. In this section, we will simulate and analyze the modulation and demodulation processes, along with the damping process, to determine the optimal number of turns. Based on this optimal outcome, we will also simulate the coherent radiation associated with various ADM processes to demonstrate the radiation performance.

Note that in the following analysis, we consider only the single-bunch dynamics. For multi-bunch operation, collective effects must be taken into account, which is beyond the scope of this work. We believe that for burst mode, a single-bunch analysis is sufficient.

\begin{figure*}
    \centering
    \includegraphics[width=1\linewidth]{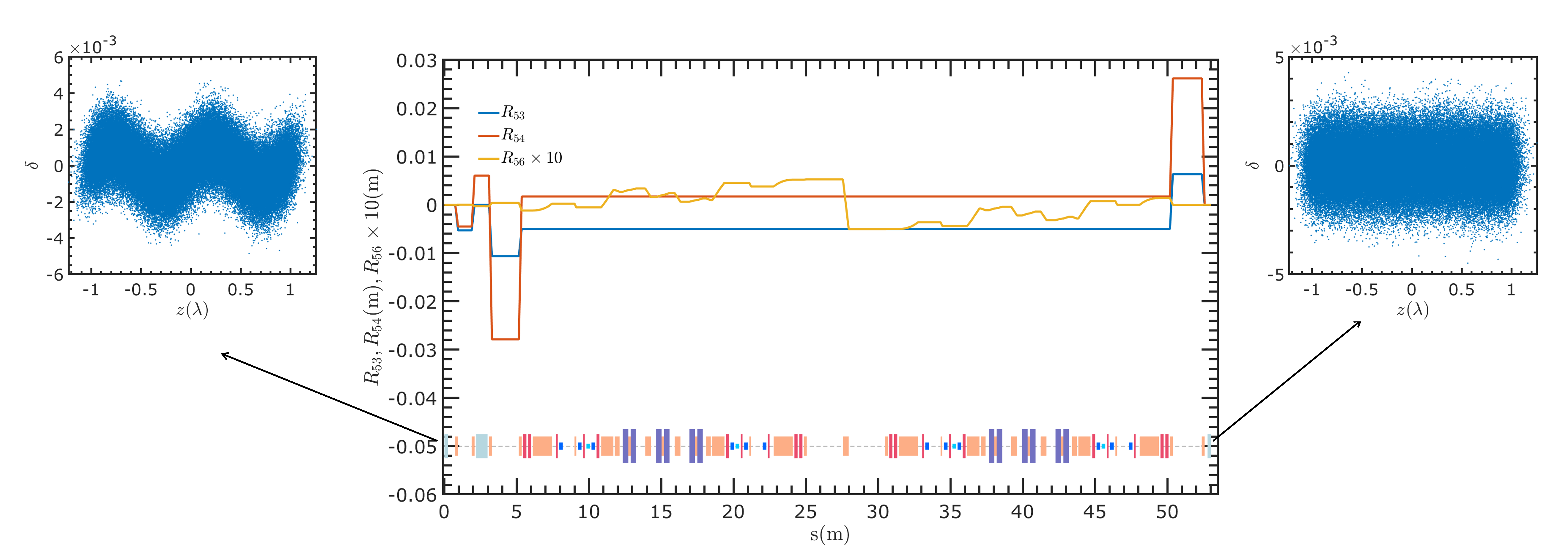}
    \caption{Transfer matrix terms $R_{53}$, $R_{54}$ and $R_{56}$ along the beam line between the center of modulator and demodulator.}
    \label{fig:6}
\end{figure*}

In the simulation of the ADM process, we employed element-by-element particle tracking in a 3D framework, as described in the previous section, to accurately capture the changes in beam quality after demodulation. However, during the damping process, both the modulation and demodulation lasers are turned off, so we can focus solely on parameter variations at the macrobunch length scale rather than the details at the microbunch scale. To optimize computational efficiency, we simplify the simulation of the damping process as follows.

We track the 4D state coordinates $(x, x^{\prime}, y, y^{\prime})$ of particles in the ring from turn $i$ to turn $i+1$ using the following one-turn map:

\begin{widetext}
\begin{equation}
\left(\begin{array}{l}
x_{i+1} \\
x_{i+1}^{\prime} \\
y_{i+1} \\
y_{i+1}^{\prime}
\end{array}\right)=\left[\begin{array}{cccc}
\cos \phi_x+\alpha_x \sin \phi_x & \beta_x \sin \phi_x & 0 & 0 \\
-\gamma_x \sin \phi_x & \cos \phi_x-\alpha_x \sin \phi_x & 0 & 0 \\
0 & 0 & \cos \phi_y+\alpha_y \sin \phi_y & \beta_y \sin \phi_y \\
0 & 0 & -\gamma_y \sin \phi_y & \cos \phi_y-\alpha_y \sin \phi_y
\end{array}\right]\left(\begin{array}{c}
x_i \\
x_i^{\prime} \\
y_i \\
y_i^{\prime}
\end{array}\right),
 \label{eq:6}
\end{equation}
\end{widetext}
where $\alpha_{x,y}$, $\beta_{x,y}$, and $\gamma_{x,y}$ are the Twiss parameters, and $\phi_{x,y}$ are the one-turn phase advances, equal to $2\pi$ times the working points $\mu_{x,y}$.

The longitudinal state coordinates $(z, \delta)$ vary as follows:
\begin{equation}
\left\{\begin{array}{l}
\delta_{i+1}=\delta_i+\frac{1}{E_0}\left[e V_c \sin \left(k_{\mathrm{rf}} z_i+\varphi_s\right)-U_0 \right]\\
z_{i+1}=z_i-C_L \alpha_p \delta_{i+1}
\end{array}\right.,
\end{equation}
where $k_{\mathrm{rf}}$ is the wave number of the RF cavity, $C_L$ is the circumference of the storage ring, $\alpha_p$ is the momentum compaction factor, $E_0$ is the energy of the storage ring, $V_c$ and $\phi_s$ are the RF voltage and phase, respectively, and $U_0$ is the one-turn energy loss.

To simulate the effects of synchrotron radiation damping and quantum excitation, we adopt Hirata's method \cite{hirata_bbc_1997}, which can be expressed as follows \cite{zhang_simulation_2005,li_cetasim_2024}. First, we transform the state vector $\mathbf{x} = (x, x^{\prime}, y, y^{\prime}, z, \delta)$ into a normalized state vector $\mathbf{X}$:
\begin{equation}
\mathbf{X} =\mathbf{M} \mathbf{x}.
\end{equation}
The normalization matrix $\mathbf{M}$ is expressed as:
\begin{equation}
\mathbf{M} =\left[\begin{array}{cccccc}
\frac{1}{\sqrt{\beta_x}} & 0 & 0 & 0 & 0 & 0 \\
\frac{\alpha_x}{\sqrt{\beta_x}} & \sqrt{\beta_x} & 0 & 0 & 0 & 0 \\
0 & 0 & \frac{1}{\sqrt{\beta_y}} & 0 & 0 & 0 \\
0 & 0 & \frac{\alpha_y}{\sqrt{\beta_y}} & \sqrt{\beta_y} & 0 & 0 \\
0 & 0 & 0 & 0 & \frac{1}{\sqrt{\beta_z}} & 0 \\
0 & 0 & 0 & 0 & \frac{\alpha_z}{\sqrt{\beta_z}} & \sqrt{\beta_z}
\end{array}\right],
\end{equation}
where $\alpha_z$ and $\beta_z$ are the longitudinal Twiss functions. Without loss of generality, we can assume $\alpha_z = 0$ and define $\beta_z$ as the bunch length divided by the energy spread $\sigma_z/\sigma_{\delta}$.

Thereafter, the effects of synchrotron radiation damping and quantum excitation are simulated using the one-turn map of the normalized vector $\mathbf{X}$:
\begin{equation}
\begin{aligned}
& \binom{X_1}{X_2}=\lambda_x\binom{X_1}{X_2}+\sqrt{\epsilon_x\left(1-\lambda_x^2\right)}\binom{r_1}{r_2} \\
& \binom{X_3}{X_4}=\lambda_y\binom{X_3}{X_4}+\sqrt{\epsilon_y\left(1-\lambda_y^2\right)}\binom{r_3}{r_4} \\
& \binom{X_5}{X_6}=\left(\begin{array}{cc}
1 & 0 \\
0 & \lambda_z^2
\end{array}\right)\binom{X_5}{X_6}+\sqrt{\epsilon_z\left(1-\lambda_z^4\right)}\binom{0}{r_6}
\end{aligned}.
\end{equation}
Here, the vector $\mathbf{r} = (r_1, r_2, r_3, r_4, 0, r_6)$ represents a five-dimensional independent Gaussian random variable. The damping coefficients $\lambda_{x,y,z} = \exp(-1/\tau_{x,y,z})$, where $\tau_{x,y,z}$ are the synchrotron radiation damping times measured in terms of the revolution time. The parameters $\epsilon_{x,y,z}$ denote the equilibrium emittances, with $\epsilon_z=\sigma_z\sigma_{\delta}$.

Finally, the state vector $\mathbf{x}$ after damping and quantum excitation can be obtained by applying the inverse normalization matrix:
\begin{equation}
\mathbf{x} =\mathbf{M}^{-1} \mathbf{X}.
\end{equation}

To develop a comprehensive understanding of beam parameter evolution as a function of the number of turns $\Delta T$, we set $\Delta T= 100$ as an example, indicating that the ADM process is conducted after every 100 damping turns. The resulting variations in beam parameters are illustrated in Fig. 7. These results demonstrate that the vertical emittance stabilizes at approximately 22 pm after around 20,000 turns, with periodic increases occurring at each 100-turn interval. The growth rate of emittance following demodulation reaches stability after approximately 100 ADM process cycles (around 10,000 turns). In contrast, the energy spread remains largely unaffected by the demodulation process, with variations primarily driven by longitudinal motion and quantum excitation, leading to fluctuations of less than 0.4\%. Given the negligible impact of these fluctuations in energy spread on the ADM process, we will focus exclusively on presenting and analyzing the changes in vertical emittance in the subsequent analysis, while acknowledging that variations in energy spread are included in the calculations but will not be explicitly discussed.
\begin{figure*}
    \centering
    \includegraphics[width=0.9\linewidth]{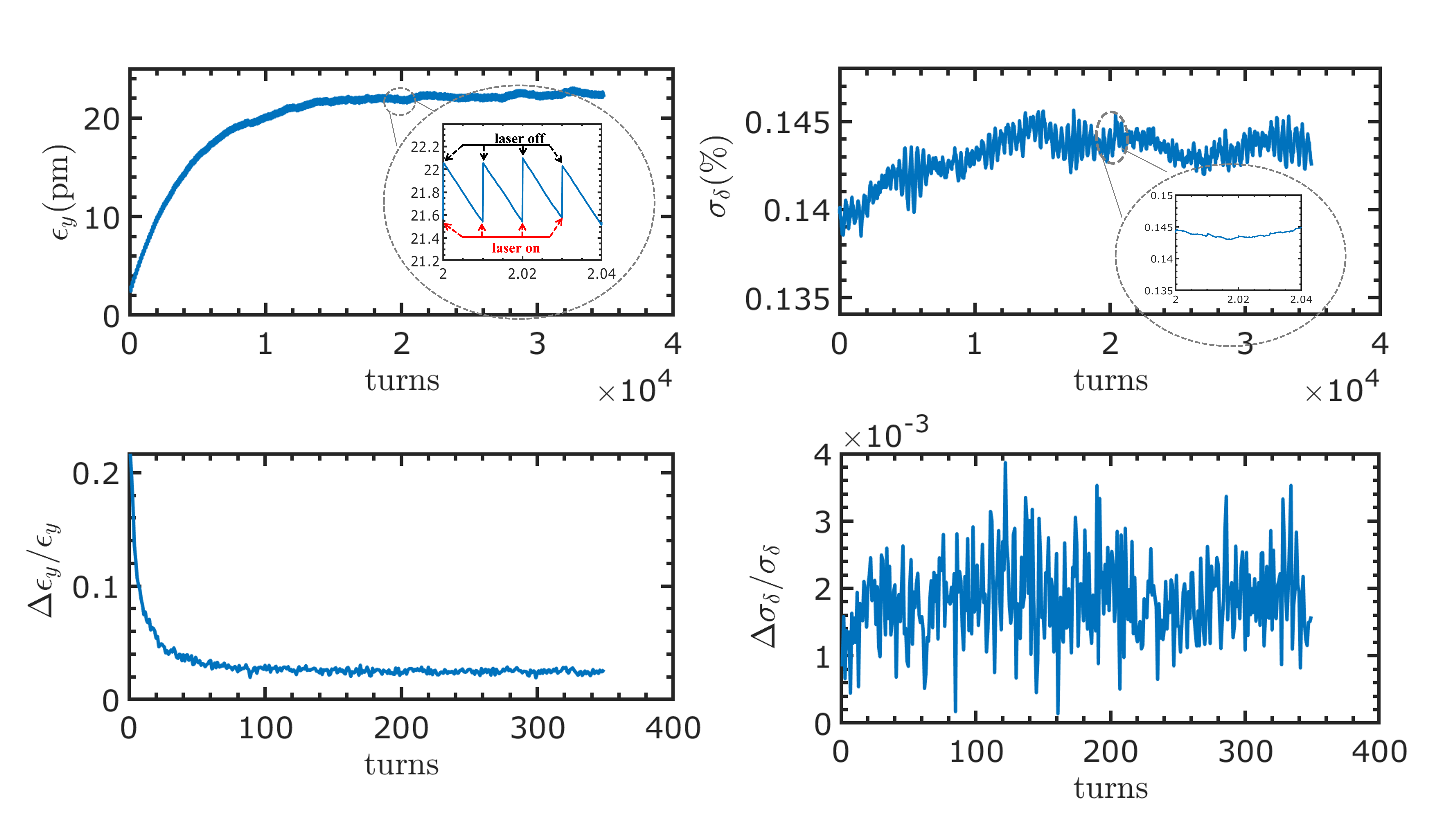}
    \caption{Variation in vertical emittance and energy spread with turns during damping and ADM processes, alongside corresponding growth rates following each demodulation.}
    \label{fig:7}
\end{figure*}

To determine the optimal number of turns $\Delta T$, we conducted a detailed analysis of vertical emittance variations across different $\Delta T$. The findings are presented in Fig. 8, which illustrates a decrease in stable emittance as $\Delta T$ increases. Notably, when $\Delta T$ is below 300, the emittance exhibits considerable variation; however, for $\Delta T$ exceeding 300, the changes become relatively minor. By examining the relationship between the bunching factor and its square (which is proportional to radiation power) as a function of $\Delta T$, it becomes apparent that further increasing $\Delta T$ beyond 600 yields only marginal improvements in radiation power.

To benchmark these results, we employed the theoretical formula for emittance variation over time under laser-off conditions \cite{DIMTR060702}:
\begin{equation}
\epsilon_y(t)=\epsilon_y(0) e^{-2 t / \tau} \tau_y+\epsilon_{y0}\left(1-e^{-2 t / \tau_y}\right).
\end{equation}
Here, $\epsilon_y(0)$ represents the emittance after demodulation obtained from ELEGANT simulations of the ADM process, while $\epsilon_{y0}$ denotes the equilibrium emittance in the vertical direction, which is 2 pm.

Using Eq. (13), we calculated the emittance after $\Delta T$ turns. The results, as shown in Fig. 8, demonstrate a good agreement between the theoretical formula and the simulation results.
\begin{figure}
    \centering
    \includegraphics[width=1\linewidth]{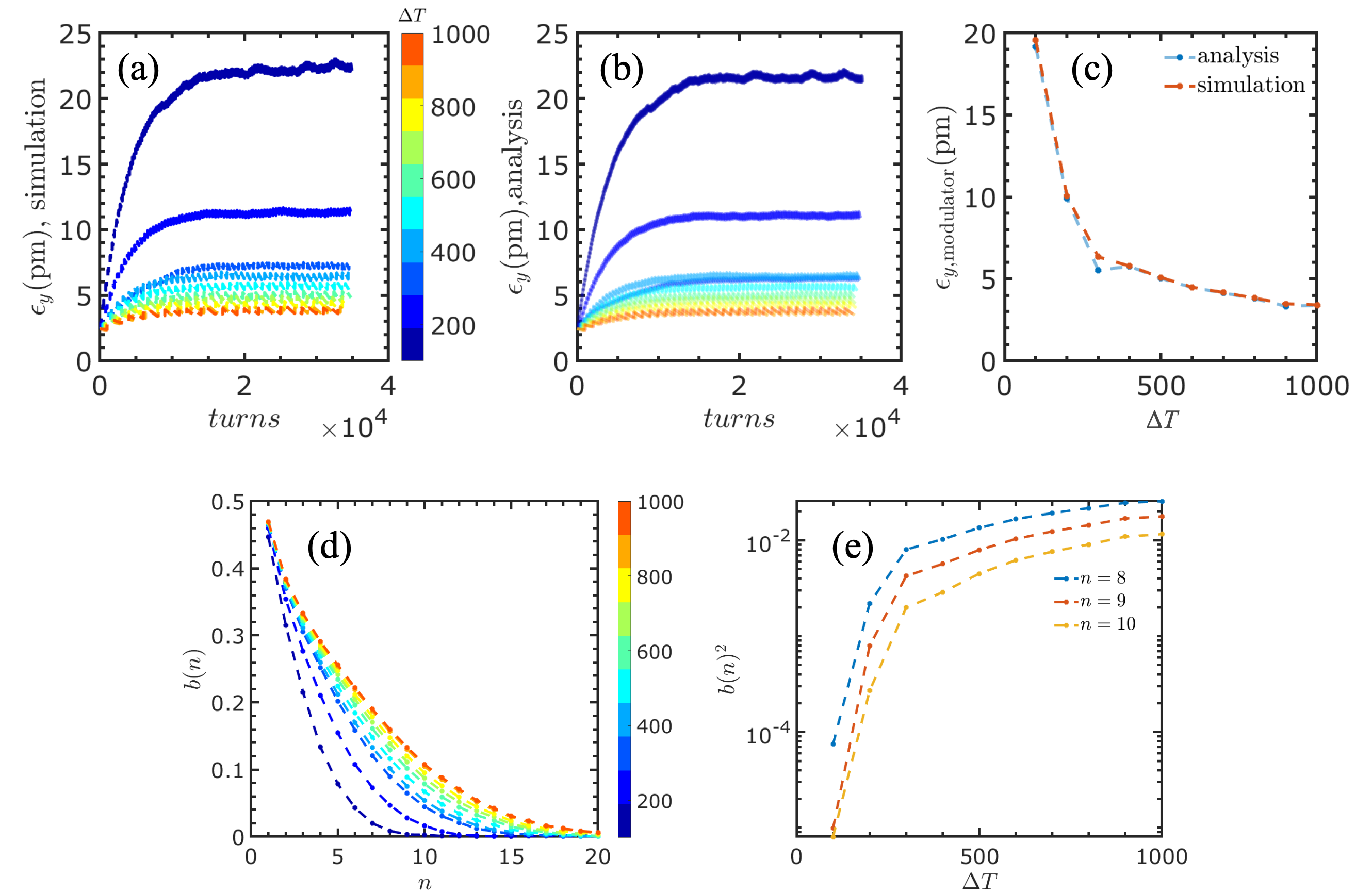}
    \caption{Variation of vertical emittance with turns under different $\Delta T$: (a) simulation results, (b) analytical results. (c) Stable emittance at the ADM modulator entrance under different $\Delta T$, (d) bunching factor at approximately 35,000 turns across different $\Delta T$ and harmonic numbers, and (e) variation of the squared bunching factor with $\Delta T$ at the target harmonic number. It is noteworthy that the fluctuations observed in the curves of (a) and (b) exhibit similar patterns, which is likely due to the use of the simulated results from each demodulation as the initial parameters for the emittance calculations using Eq. (13).}
    \label{fig:8}
\end{figure}

We take the optimal $\Delta T$ as 600, meaning that the ADM process is performed approximately every 1.62 ms. Considering that the storage ring contains 450 buckets with a 90\% filling pattern, the SAPS reversible ADM can achieve a maximum repetition rate of approximately 0.25 MHz.

Notably, after the beam stabilizes around 11,000 turns, small differences in beam parameters persist at the modulation entrance, despite the overall stability. These discrepancies can lead to variations in the bunching factor, which in turn affect radiation performance. To assess the significance of these differences, we calculated the radiations produced during the ADM process from 11,000 to 35,000 turns. Beginning at 11,000 turns, we activated the laser every 600 turns when the beam reached the ADM section. Following modulation with a laser of 700 fs duration, the beam passing through the radiator generated coherent radiation. A total of 40 radiations were simulated using the GENESIS code \cite{reiche_genesis_1999}, with each radiation pulse generated every 600 turns. 

\begin{figure*}
    \centering
    \includegraphics[width=0.9\linewidth]{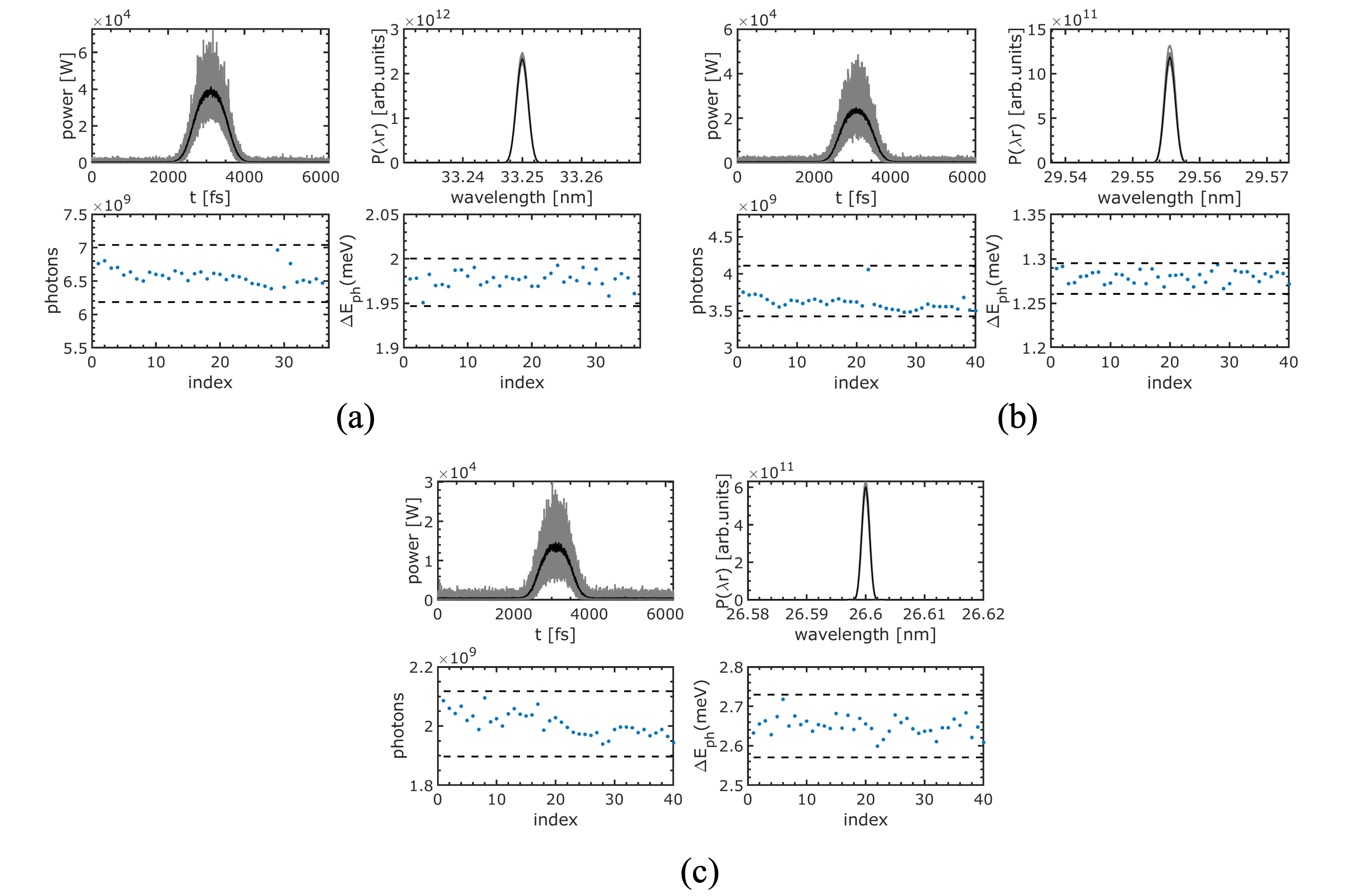}
    \caption{Radiation pulse and spectrum for different ADM processes, with the black solid line representing the average result. Also shown are the pulse-by-pulse variations in photon number and bandwidth. Panels (a) to (c) represent different radiation wavelengths.}
    \label{fig:9}
\end{figure*}

As shown in Fig. 9, the duration of these radiation pulses is approximately 1 ps and the photon numbers per pulse is approaching $7\times10^{9}$ for radiation pulse with wavelength of 33.25 nm, exhibiting an approximately 6\% variation among different pulses. Considering a repetition rate of 0.25 MHz, the average flux reaches $1.8\times10^{15}$ photons/s. 
The bandwidth of the radiation pulses is approximately 2 meV, with variations remaining below 1.3\%. The performance for other radiation wavelengths is also shown in Fig. 9.

For comparison, we estimated the photon flux produced by three 6-meter-long undulators under equivalent bandwidth conditions to be around $2.5\times10^{13}$ photons/s. Although this estimation may be higher than the actual photon flux, it still falls beyond one orders of magnitude lower than that of coherent radiation.

\section{Discussion and conclusion}

We propose an easy-to-implement method to achieve reversible laser modulation in storage rings. This scheme utilizes two arcs and three straight sections of the storage ring. By incorporating a chicane within one of the straight sections, an isochronous transfer from the modulation section to the demodulation section is achieved, thereby eliminating energy modulation. Based on SAPS parameters, we present a design. The results indicate that after beam modulation, a large bunching factor can be achieved at the high harmonic. Under modulation by a laser with several hundred femtoseconds duration, picoseconds radiation pulse can be achieved with large average flux.

After demodulation, the beam parameters are largely recovered. However, compared to the case under ideal conditions, the beam energy spread and vertical emittance increase significantly. This difference is attributed to higher-order terms in the transport. Future efforts will focus on reducing this discrepancy by locally optimizing the nonlinear elements in the arcs. At the current growth rate, beam recovery requires approximately 1.62 microseconds, allowing for a repetition rate of up to 0.25 MHz.

It is noteworthy that the reversible ADM has some effects on the storage ring dynamics. Since the longitudinal dispersion of the transport between modulation and demodulation is zero, the momentum compaction factor of the whole ring will be reduced. This reduction will have some effect on the longitudinal dynamics, although in our case the momentum compaction factor is only reduced by $2/32 \sim 6$\% due to the multiple periods of the SAPS, and the effect on the longitudinal dynamics is negligible. For storage rings with a few periods, e.g., only two periods, this effect is significant. For this kind of storage ring design, steady-state microbunching storage ring related studies can be used as a guide \cite{deng_experimental_2021,deng_steady-state_2024}. 

In addition, the scheme utilizes three straight sections and two arcs, which presents a clear disadvantage in terms of beamline utilization. This issue could be addressed by designing a storage ring with longer straight sections, thereby enabling the scheme to be implemented with only two straight sections. Alternatively, longitudinal dispersion of the chicane could be achieved by utilizing the horizontal components of the dipoles in the ADM and demodulation sections, although this approach may increase the complexity of mechanical alignment. Further research into more effective solutions to this challenge is warranted.

Nevertheless, in this study, we have not addressed the specific implementation of laser-beam synchronization, assuming perfect synchronization. Currently, synchronization between the storage ring beam and the laser can achieve jitter on the order of a few ps \cite{TANAKA2000}. For a storage ring bunch length of several hundred ps, this level of jitter is sufficient. However, for shorter bunch lengths on the order of a few ps, we aim to reduce jitter to the tens of fs level. This is potentially achievable, as fs-level jitter has already been demonstrated in FELs \cite{Danailov2014}. Future research that integrates FEL-related technologies for laser-beam synchronization in storage rings could help achieve this goal.

\begin{acknowledgments}
This work was supported by National Natural Science Foundation of China (Nos. 12405176, 12275284 and 11922512), and Youth Innovation Promotion Association of the Chinese Academy of Sciences (No. Y201904).

\end{acknowledgments}

\appendix
\onecolumngrid 
\section{Parameters table of the Reversible ADM section}
For convenience, we have compiled the parameters used throughout this paper in Table I below.

\begin{table}
  \centering
  \caption{Parameters of the Reversible ADM section}
    \begin{tabular}{lll}
    \toprule
    Parameter & Value & Unit \\
    \midrule
    \multicolumn{3}{c}{Ring} \\
    \midrule
    $C_L$ & 810 & m \\
    $\alpha_p$ & $2.334\times10^{-5}$ & \\
    $U_0$ & 0.921 & MeV \\
    $V_c$ & 3 & MV \\
    $\mu_x,\mu_y$ & (7.6123, 4.9091) &  \\
    $\beta_x,\beta_y$ & (6.75, 6.41) & m \\
    $\alpha_x,\alpha_y$ & (0.613, 0.6657) & \\
    $k_{\mathrm{rf}}$ & 3.49& 1/m \\
    $R_{56,r}$ & 0.638 & mm \\
    Damping time ($t_x$, $t_y$, $t_{\delta}$) & 13.7, 20.5, 13.6 & ms \\
    Peak current & 31.5 & A \\
    \midrule
    \multicolumn{3}{c}{ADM section} \\
    \midrule
    Dipole & & \\
    \midrule
    Bending angle & 5 & mrad \\
    Length & 0.20 & m \\
    \midrule
    Dogleg & & \\
    \midrule
    Bending angle & 5.31 & mrad \\
    Length & 0.2 & m \\
    Distance between dipoles & 0.94 & m \\
    \midrule
    Modulator & & \\
    \midrule
    Peak field & 0.6 & T \\
    Period & 0.25 & m \\
    Period number & 1 & \\
    \midrule
    Radiator & & \\
    \midrule
    Peak field & 0.54 & T \\
    Period & 0.12 & m \\
    Period number & 7 & \\
    \midrule
    \multicolumn{3}{c}{Laser} \\
    \midrule
    Peak power & 62.4 & GW \\
    Waist size & 0.14 & mm \\
    Duration & 700 & fs \\
    \midrule
    \multicolumn{3}{c}{Chicane} \\
    \midrule
    Dipole bending angle & 13.91 & mrad \\
    Dipole length & 0.2 & m \\
    Distance between first and second dipoles & 2.5 & m \\
    \midrule
    \multicolumn{3}{c}{Demodulation section} \\
    \midrule
    First dipole length & 0.2 & m \\
    First dipole angle & 9.59 & mrad \\
    Second dipole length & 0.2 & m \\
    Second dipole angle & -6.73 & mrad \\
    Distance between first and second dipole & 2 & m \\
    Third dipole length & 0.2 & m \\
    Third dipole angle & -11.02 & mrad \\
    Distance between second and third dipole & 0.4 & m \\
    Fourth dipole length & 0.2 & m \\
    Fourth dipole angle & 8.15 & mrad \\
    Distance between third and fourth dipole & 2.6 & m \\
    \bottomrule
    \end{tabular}%
  \label{tab:A}%
\end{table}%

\twocolumngrid %


\nocite{*}

\bibliography{paper}% Produces the bibliography via BibTeX.

\end{document}